# Employees' Productivity Measurement and Control – A Case of a National University

Khaled Waleed
Senior IT Auditor
University of Bahrain
Kingdom of Bahrain
+973-17437753
Kalamawieh@uob.edu.bh

Ali AlSoufi
Associate Professor
British University of Bahrain
Kingdom of Bahrain
+973-17621501
Ali.AlSoufi@Gmail.Com

## ABSTRACT
An experimental study, that finds the impact of Internet access control on the employees' productivity in the National University. The purpose of the study is to boost the employee productivity through proper Internet access control. The main objectives are to find the most used web categories by the staff, find if relation exists between productivity and non-work-related internet usage, and choose the best level of Internet access control. Before initiating the experiment, Employees' Internet usage was monitored and accordingly classified them into the proper Internet access control groups. Then supervisors were asked for a pre-test productivity measures for their staff, after that the experiment was initiated for 45 days. Then, a post-test productivity measure was done. Productivity changes were analyzed with the department nature, its Internet usage portfolio and its current Internet access control group; then the best level of restriction was found. The result showed that the productivity of departments with low Internet usage was not affected by restricting/unrestricting Internet access. However, for high Internet usage departments noticeable productivity improvement was there when the Internet restriction policy was not affecting work-related websites; but when it was affecting work-related websites the productivity decreased.

## 1. INTRODUCTION
### 1.1 Background of the Study
"As the 21st century progresses, the Internet is showing no signs of slowing in growth" (Kimberly & Carle 2002). Each year the number of Internet users increases, "in 2013 2.7 billion people – almost 40% of the world's population – is online" (ITU, 2013). As the IT and communication technologies advances, the Internet becomes critical to business operation. No question that most organizations can't operate without the Internet, actually the percentage of employees in companies having Internet access is growing every year, according to Kimberly & Carle (2003), a human resource director's survey showed that "approximately 70% of companies provide Internet access to more than half of their employees". A recent survey showed that 82% of computer-using employees engaged in non-work-related Internet browsing during work hours" (Anandarajan, Simmers & D'Ovidio, 2011). Kathleen (2012) stated that Sex-Tracker reports that 70% of all Internet porn traffic occurs during the 9-to-5 workday she also added that such workplace Internet misuse costs U.S. companies $63 billion in lost productivity annually, according to Websense Inc.

No doubt that Internet access improves the productivity of the firm, Grimes & Stevens (2012) found that the boost in productivity is between 7% to 10%. However many organizations –as mentioned above- complain that the Internet is also causing a loss in productivity since it staff are misusing it, therefore most of the organizations started applying Internet access control techniques.

### 1.2 Objectives of the Study
The purpose of the research is to improve the employees' productivity through Internet access control. The following are the objectives of the study:

1. To determine web categories that utilizes most of the staff time.
2. To find the relation between employee productivity and non-work-related Internet usage.
3. To determine the best level of control that will boost the productivity (un-restriction, semi-restriction or full-restriction).

## 2. LITRETURE REVIEW
Each year the number of Internet user's increases, in 2013 2.7 billion people – almost 40% of the world's population – is online (ITU, 2013). The Internet became an essential part of the business. Many of business and jobs require the use of modern technology in their daily operation. Internet boosts the companies' productivity, which, in turn increases the revenue. "Results indicate that broadband adoption boosts firm productivity by 7–10%" (Grimes & Stevens, 2012).

More than that, Internet will also give a competitive edge to the business and increase its market selection. Therefore, most of today's companies and governmental organizations are widely providing Internet access to their employees. According to Kimberly & Carle (2003), a human resource director's survey showed that approximately 70% of companies provide Internet access to more than half of their employees.

### 2.1 Negative effects of Internet Misuse to Companies
- **Productivity Loss**: This productivity loss is translated into large financial figures. According to Young (2011), Internet abuse costs billions in lost productivity annually.
- **Technical Costs** "Internet access costs money, either in fees to support servers, Internet Service Providers, or in hardware costs necessary to accommodate increased network traffic and data storage." (Young, 2011).
- **Social and Personal Drawbacks:** Internet abuse affects the social life in the work place by "creating new social problems

in the workplace when employees become socially withdrawn, dislike real life meetings, working in collaborative teams and only prefer online communication, fearing face-to-face contact" (Young, 2012).

## 2.2 Positives of Internet Access Restriction

The main reason companies restrict Internet access is to increase productivity. Riedy & Wen (2010) summarized the most critical issues that makes Internet access restriction a must:
- Potential corporate liability for employee misconduct via the Internet.
- Diminished employee productivity from recreational use of the Internet
- The unauthorized disclosure of proprietary information and viruses attacks.

## 2.3 Negatives of Internet Access Restriction

The subjects of the researchers on the negatives of Internet access control is mainly divided into the following sections:
1. Technical and legal issues related to the monitoring techniques and their use.
2. Financial and managerial issues related to the cost of these appliances and the negative effects they will have on employees from a managerial perspective.
3. Social and physiological issues related to the privacy, trust of the organization and relation between the manager and the employee

## 2.4 Researchers' Results

According (Anandarajan et al., 2011) the literature on PWU in the workplace grew substantially over the last few years in three streams:
1. The first stream is problematic Internet use (PIU). PIU is a multidimensional syndrome consisting of cognitive and behavioral symptoms resulting in negative social, academic, or professional consequences. The emphasis in PIU is on the association between Internet use and psychosocial health. In this view, Internet abuse is often assessed as completely negative for both employer and employee.
2. A second stream examines Internet abuse as a variation of production deviance. Production deviance pertains to violations of formally prescribed norms in the workplace. In the Internet abuse literature these behaviors are typically referred to as cyber loafing.
3. A third view and less common stream in research is viewing personal web usage as a constructive behavior. While organizations often overlook and seldom use to this to their advantage, in practice, employees recognize the positive benefits of personal web usage".

However, most of the studies look to the Internet abuse and productivity loss issue from a managerial perspective only without considering the technical part. Therefore, most of the studies do not conduct an experimental environment to control the Internet access restriction nor do they consider the level of restriction applied, blocked categories, bandwidth allocated for the allowed categories, types of proxies and monitoring software applied and the Internet usage reports. Furthermore, studies do not measure the employees' productivity formally by contacting their supervisors; instead, they will count on surveys where each employee measures his own productivity. Therefore, the results are inaccurate and incomplete and this study came to full-fill these missing parts.

# 3. RESEARCH METHODOLOGY
## 3.1 Research Design

There are two research design methods that will be used: Primary data and evaluation method.

Figure 3.1 shows the research diagram that summarizes the major steps followed in the research

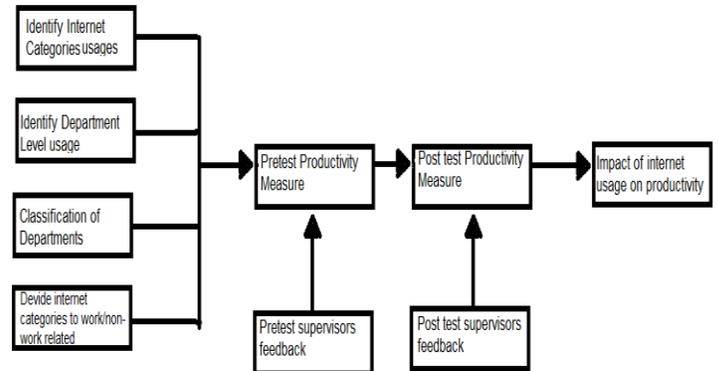

**Figure 3.1: Research Diagram**

There are two research design methods that will be used: Primary data and evaluation method. The main primary data sources for employee Internet usage were obtained from special devices for Internet control, monitoring and reporting; this package is known as Bluecoat Reporter and Bluecoat SG9000 proxy. The primary data will help to:
- Determine the current Internet utilization of each department
- Compare the usage of the different web categories
- Determine the mostly used web categories
- Identify the classification of departments into restricted and unrestricted groups.

For the evaluation, method there will be two activities:
- **Pre-test staff productivity evaluation**: This will measure the employee productivity before changing the Internet access. Each director is responsible to measure his staff productivity
- **Post-test staff productivity evaluation:** After one month of applying the restrictions/un-restrictions in the Internet policies for the staff, the directors and managers have to evaluate their staff again.

## 3.2 Study Population

The sample size is 161 staff. However, the target population of this research is the administrative staff of the University. In fact, that staffs are divided into **ten** departments. Technically speaking, these departments are divided into departments falling under a single dedicated computer virtual LAN (VLAN) for staff only or shared computer VLAN where staff and student will be in it.

## 3.3 Sampling Procedure

The following procedures are followed to choose the sample frame:
1. Determine the target population: University of Bahrain Administrative staff.
2. Choose Non-Probability Design as Sampling Design.
3. Choose the Convenience Sampling techniques which means the sampling procedure of obtaining the people or units most conveniently available.

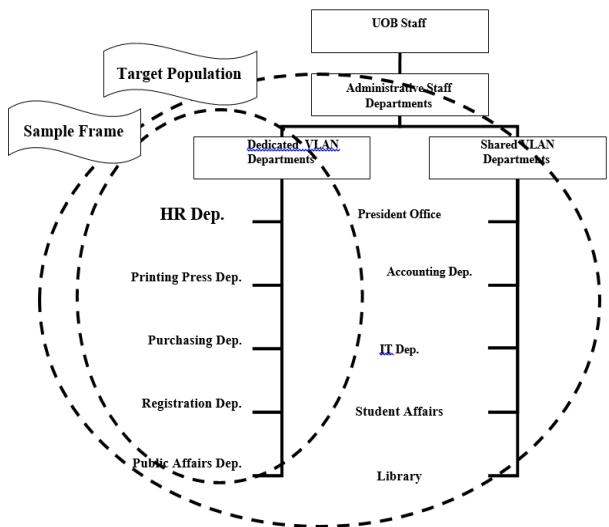

**Figure 3.2: Sample Frame for the Research**

After sampling, the research will have 161 administrative staff. As shown in figure 3.3, the reason I excluded the Academic staff (UOB collages) from the research because the majority of staff are in the teaching field and they are only setting on their desk for a very limited time. The reason for exclusion of departments, which have shared VLANs from the Sample frame, is for technical reasons: since the device that monitors Internet usage in the different departments cannot differentiate staff from students in shared VLANs. This will make some parts in the research almost ambiguous since the data will not be tagged with the user address or the user type: Student or Administrative Staff or Academic Staff.

## 3.4 Instrumentation

Several data collection sheets were needed to collect the data that is necessary to answer the research questions and achieve the research objectives. The data collection sheets are divided into four main parts:

**Instrument 1:** Data collection sheets to collect information about departments' current Internet usage in top used web categories. It will determine the amount of utilization of the staff for each web category. In order to do this classification of departments into restricted and non-restricted, the following steps are to be done:
1. Find out the top used categories in the study population by generating the report form Bluecoat Reporter.
2. Classify these categories into work related categories and non-work-related categories.
3. Find out the Internet utilization of each department with respect to the top used categories and fill-up the data collection sheet.
4. Based on these results classify the departments into restricted Internet access and unrestricted Internet access groups.

Below is the top ten categories used in the UoB according to Bluecoat Proxy SG and Bluecoat Reporter website classification with a description of each web category:

o **Audio video clips:** Sites that provide streams or downloads of audio or video clips-typically.
o **Open/Mixed Content:** Sites with generally non-offensive content but that also have potentially objectionable content such as adult or pornographic material that is not organized so that it can be classified separately.
o **Chat/Instant Messaging:** Sites that provide chat, text messaging (SMS) or instant messaging capabilities or client downloads.
o **Social Networking:** Sites that enable people to connect with others to form an online community
o **Search Engines/Portals:** Sites that support searching the Internet, indices, and directories.
o **Computer/Internet:** Sites that sponsor or provide information on computers, technology, the Internet and technology-related organizations and companies.
o **News/Media:** Sites that primarily report information or comments on current events or contemporary issues of the day.
o **Forums/Blogs:** Sites that primarily offer access to personal pages and blogs.
o **Others:** not all the other categories listed above.

- **Instrument 2:** Data collection sheets to collect the data that are related to evaluate the Staff Productivity before the test.

**The pre-test evaluation Process**
Envelops will be prepared for each head of section in-order to evaluate his staff in an anonymous way and then send back the results to the IT Center. Each envelop for the heads of sections contains the following: Explanatory letter, Staff Productivity Evaluation form and Staff Key Code. The following are steps was followed by the head of sections in order to complete the evaluation process:
1- Open the white envelope and fill-up the Staff Code Key form by putting a number of your choice next to each staff name. Now this number will be considered the staff ID for the experiment
2- Fill-up the Staff Productivity Measurement form for each employee under your supervision. On the employee ID section type the number, you had chosen in the Staff Code key sheet without mentioning the employee's name to preserve the privacy of the evaluation.
3- Put the Staff Code Key in the white envelope and close it - with the seal of the department - and place the white envelope in the brown envelope (A4) with the evaluation forms. Then send it to the Information Technology Center.
4- After 40 days, the Information Technology Centre will return the sealed white envelope to you without opening it with the new evaluation forms.

- **Instrument 3:** Data collection sheet to calculate the speed of non-work-related web categories for all access control groups. The purpose of this instrument is to show the difference in time for access restricted websites and unrestricted websites and prove if the restricted access policy on Bluecoat Proxy SG is really functioning properly.

- **Instrument 4:** Data collection sheets to collect the data that are related to evaluate the Staff Productivity after the test is finished.
**The post-test evaluation Process**
The following steps is to be followed by the head of sections in order to complete the evaluation process:
1. Fill-up the Staff Productivity Measurement form for each employee under your supervision. On the employee ID section type the number, you had chosen in the Staff Code key sheet without mentioning the employee's name to preserve the privacy of the evaluation.

2. Your evaluation of your staff productivity will be for the past 40 days only.
3. Get rid of the Staff Code Key.
4. Send the Productivity Evaluation form back to the IT Center.

## 4. DATA ANALYSIS AND RESULTS

I organized this chapter by dividing the chapter into several sections according to the research questions. The main points are:
1. Web categories that utilizes most of the employees browsing time
2. Relation between employee productivity and non-work-related internet usage
3. The best level of web access control to boost employee productivity.

### 4.1 Employee Browsing Time for Each Category

This graph will show the average University administrative staff utilization for the top web categories. It can be seen from figure 4.1 how many minutes spent on each web category on a working day for a single user.

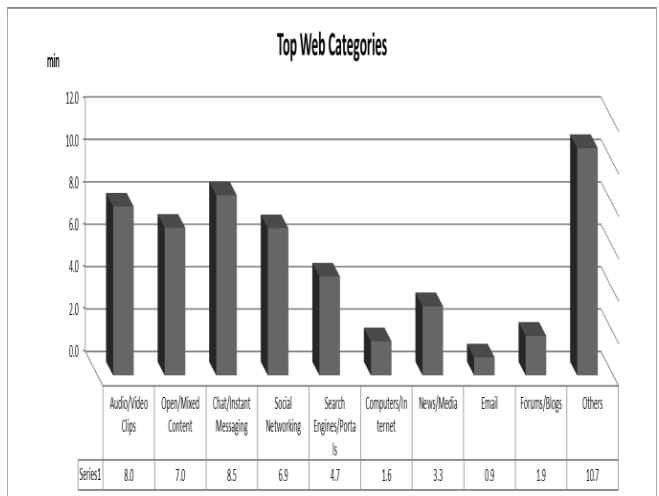

**Figure 4.1: Top Web Categories Utilization**

### 4.2 Relation between Productivity and Non-Work-Related Internet Usage

This section is to answer the statement: Is there a relation between employee productivity and non-work-related internet usage? That means weather misusing the Internet at work affect the staff productivity or not.
Table 4.1 shows the pre-test productivity for each department:

**Table 4.1: Average productivity for each department**

| Department | Pre-test |
|---|---|
| Public Affairs | 86.00% |
| Human Resource | 93.75% |
| Purchasing | 86.67% |
| Registration | 95.71% |
| Printing Press | 100.00% |

### 4.3 Total Web Usage and Non-Work Categories

When comparing the total browsing time and the non-work-related categories browsing time for the five departments as a percentage out of hundred, we found that they are almost equal refer table 4.2 for details. This means that high web users will spend long periods in non-work-related internet related categories and low web users will spend little periods in non-work-related categories.

| Department | Non-work categories browsing time per user (min) | Non-work categories utilization | Total browsing time per user (min) | Total categories utilization |
|---|---|---|---|---|
| Public Affairs | 52.2 | 37% | 73.3 | 37% |
| Human Resource | 32.2 | 23% | 50.8 | 24% |
| Purchasing | 20.3 | 14% | 33.2 | 16% |
| Registration | 20.0 | 14% | 31.4 | 15% |
| Printing Press | 17.3 | 12% | 24.6 | 12% |

**Table 4.2: Average utilization of non-work categories**

### 4.4 Results

Figure 4.2 shows the relationship between non-work-related browsing and productivity a line chart was drawn. From the chart, it can be seen that there is almost a linear relation between the non-work-related Internet browsing and the productivity. Hence when the non-work browsing time increases productivity decreases.

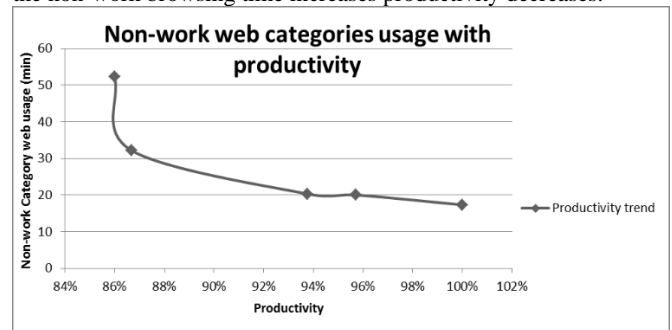

**Figure 4.2: Non-work-related internet categories usage against productivity**

### 4.5 Departments' Utilization per Web Category

Using more advanced quires on the Bluecoat Reporter, more detailed reports about each department was generated. The bar chart in figure 4.3 will give an idea how much time staff from each department will spend in each web category individually. It is clear that most of the top used websites categories are in the non-work related-group

## 5 Conclusion

In conclusion the research is summarized briefly in the following points:
- In the University, most of the web browsing time is spend on non-work purposes. The study showed that almost 70% of the browsing time is in non-work-related browsing
- Many of the top used web categories are the most bandwidth greedy categories consuming around 50% of the browsing bandwidth, like: Audio/Video, Social Networking, and News/Media.
- Productivity was found to be having a negative relation with non-work-related Internet usage
- Internet access control –weather restricting or unrestricting- did not affect the productivity of departments who got low

Internet utilization profile. As seen in Purchasing, Registration and Printing Press.
- Restricted web access showed a noticeable improvement in the productivity of the department which got a high utilization profile of non-work categories, given the condition that these non-work categories are not part of the department's job. As can be seen in HR.
- High web utilization department's productivity was decrease when the restricted access control policy was blocking categories that are part of the department's job as in Public Affairs. A single web access control policy will not affect the productivity evenly for all departments. Hence, internet access control should be customized depending on each department's needs.
- In the University, the web access policies on table 4.7 should be applied to boost the departments' productivity. When more than one "Yes" is there this means that all the ticked polices will have the same effect on productivity.
- Internet access control is not only limited to IT, management decisions are also part of it.
- Applying open office system is actually type of Internet access control as was seen in the purchasing department

## ACKNOWLEDGMENTS

I would like to send my gratitude to my supervisor Dr. Ali Alsoufi for his advises and kindness. In addition, I would like to thank Dr. Khalid Almutawa the head of IT Center in the University for his great efforts for the successful accomplishment of the experiment. Finally, I would like to acknowledge everyone helped me to finish the research.